# Revealing the electron-spin fluctuation coupling by photoemission in CaKFe$_4$As$_4$

Peng Li[1,2], Yuzhe Wang[1,3], Yabin Liu[4], Jianghao Yao[1,3], Zhisheng Zhao[1,3], Zhengtai Liu[5], Dawei Shen[6], Huiqian Luo[7], Guanghan Cao[4], Juan Jiang[1,3,†] and Donglai Feng[1,3,6,†]

[1]*School of Emerging Technology, University of Science and Technology of China, Hefei 230026, China*
[2]*Quantum Science Center of Guangdong–Hong Kong–Macao Greater Bay Area (Guangdong), Shenzhen 518045, China*
[3]*Hefei National Laboratory, University of Science and Technology of China, Hefei 230088, China*
[4]*Department of Physics and State Key Lab of Silicon Materials, Zhejiang University, Hangzhou 310027, China*
[5]*State Key Laboratory of Functional Materials for Informatics, Shanghai Institute of Microsystem and Information Technology (SIMIT), Chinese Academy of Sciences, Shanghai 200050, China*
[6]*National Synchrotron Radiation Laboratory, School of Nuclear Science and Technology, and New Cornerstone Science Laboratory, University of Science and Technology of China, Hefei, 230026, China*
[7]*Beijing National Laboratory for Condensed Matter Physics, Institute of Physics, Chinese Academy of Sciences, Beijing 100190, China*

**ABSTRACT**

Electron-boson coupling in unconventional superconductors is one of the key parameters in understanding the superconducting pairing symmetry. Here, we report definitive photoemission evidence of electron-spin exciton coupling in the iron-based superconductor CaKFe$_4$As$_4$, obtained via high-resolution ARPES. Our study identifies a distinct kink structure on the α band, observable only in the superconducting phase and closely linked with the superconductivity, indicative of strong electron-boson interactions. Notably, this kink structure corresponds to two distinct bosonic modes at 11 meV and 13 meV, aligning with spin resonance modes previously observed in inelastic neutron scattering experiments. This alignment underscores the significant role of antiferromagnetic fluctuations in the pairing mechanism of this superconductor. Furthermore, the unique momentum-dependent and orbital-selective properties of the coupling revealed by ARPES provide profound insights into the pairing symmetry, suggesting predominantly s$_\pm$ wave pairing facilitated by spin fluctuations. Our findings not only highlight the pivotal role of spin resonance in the superconductivity of CaKFe$_4$As$_4$ but also enhance understanding of the electron-spin exciton interactions in unconventional superconductors.

**INTRODUCTION**

Understanding the pairing mechanism in Fe-based superconductors continues to be a pivotal focus in condensed matter physics. Deviating from the phonon-mediated conventional Bardeen-Cooper-Schrieffer (BCS)-type superconductors, strong electronic couplings, such as nematicity and magnetism in particular, interplay with superconductivity in unconventional superconductors [1-7].

Antiferromagnetic fluctuations are commonly identified as the 'pairing glue' in these materials, manifested by the spin resonance modes, or the spin-1 excitons, generated by particle-hole spin excitations in the superconducting state. Indeed, numerous inelastic neutron scattering (INS) experiments have observed the spin resonance modes with energies scaling with the transition temperature ($T_c$), providing substantial evidence for magnetically mediated copper pairing in these unconventional superconductors [8-20]. In parallel, angle-resolved photoemission spectroscopy (ARPES) studies have revealed kink structures in the band dispersion, which is related to certain electron-boson couplings. Except for the electron-phonon coupling, these couplings usually include electron-antiferromagnetic magnon coupling and electron-spin fluctuation coupling. These bosonic modes can be determined through ARPES spectra self-energy analysis to extract the energy positions and widths of these modes, offering a direct comparison with INS results [21-31]. Moreover, ARPES allows for the exploration of the momentum-dependent and orbital-selective properties of these couplings, offering unique insights into the underlying pairing mechanisms. However, despite these advances, robust ARPES evidence for electron-spin fluctuation couplings in Fe-based superconductors remains elusive [24,32].

The recently discovered stoichiometric bilayer Fe-based superconductor $CaKFe_4As_4$ offers an exemplary platform for probing electron-spin exciton coupling, owing to its high superconducting transition temperature ($T_c \sim 35$ K) [33,34] and clear spin resonance modes in INS experiments [11,13,35]. Previous ARPES studies have confirmed $CaKFe_4As_4$ as a prime candidate for a topological superconductor [36], featuring orbital-dependent superconducting gaps [37]. Additional, a recent scan tunneling spectroscopy (STS) experiment suggested potential electron-boson coupling therein, with the energy of the bosonic mode aligning closely with the average resonance modes detected in INS [38]. Nevertheless, definitive evidence for electron-spin fluctuation coupling in $CaKFe_4As_4$, particularly concerning its orbital characteristics and momentum dependence, remains unexplored, which could deepen our understanding of the pairing mechanisms of Fe-based superconductors.

In this letter, we have detailed a comprehensive ARPES investigation of $CaKFe_4As_4$, which has revealed exceptional data quality and distinct kink features. The kink only appears below $T_c$ and is located on the $\alpha$ band, which is characterized by $d_{xz}/d_{yz}$ orbitals and exhibits the largest superconducting gap simultaneously. Unlike other bands which show quasi-two-dimensional characteristics, the α band exhibits bilayer-splitting near $k_z \sim \pi$. Interestingly, both the kink position and superconducting gap of the $\alpha$ band have a weak dispersion along the $k_z$ direction, corresponding to bosonic mode energies of 11 meV at $k_z \sim 0$ and 13 meV at $k_z \sim \pi$. These bosonic modes coincide with the two odd spin resonance modes (10.5 meV and 13 meV) observed in INS. Furthermore, the in-plane distribution of the superconducting gaps supports a sign-change s-wave ($s_\pm$) pairing symmetry in this material. Our findings provide direct photoemission evidences of the electron-spin resonance coupling in iron-based superconductors.

**RESULTS**

$CaKFe_4As_4$ exhibits a bilayer FeAs structure with a total self-doping level of 0.25 hole/Fe, as documented in prior studies [34,39]. Figure 1a shows the Fermi surface maps focused on Brillouin zone center (Z) and Brillouin zone corner (A) at $k_z \sim \pi$ plane using 30 eV and 60 eV photons, respectively. The maps reveal at least four hole pockets (labeled α/α1, β, and γ) around the Z point and two orthogonal electron pockets (labeled δ and ε) around the A point. Notably, bilayer splitting, marked by overlapping deep and light blue dashed lines on the innermost α pocket, is more pronounced in the *ZAR* plane. In order to realize its three-dimensional band structure, the band dispersions of these bands at $k_z = \pi$ (ZAR) and $k_z = 0$ (ΓMX) are presented in Figs.1b, 1c and 1d. The α band is degenerate at the ΓMX plane, however, at the ZAR plane a clear splitting can be observed along the ZA direction, evident from the momentum distribution curves (MDC's) [more details can be found in Fig. S1 in Supplementary Materials]. Most of the bands exhibit a quasi-two-dimensional nature with band dispersions along the high-symmetry cut showing minimal variation in the $k_z$ direction. This is evident from the fully overlapped peak positions of the α, β, γ, ε and δ bands from the MDC's near the Fermi level (Figs. 1b, c and d). One could notice that there are some residual intensities at the Z and Γ points, which have been proved to be the topological surface states [36] and is not the focus here [details can be found in Fig. S2 in Supplementary Materials]. The $k_z$ dependent behaviors of the α, β, γ bands are summarized in Fig. 1e, except for the α band which splits into α and α1 near the Z point, all other bands exhibit quasi-two-dimensional behaviors.

The $k_z$ dependence of superconducting gaps across various bands, as illustrated in Figs. 2a and 2b, demonstrates notable variations in gap magnitude as evidenced through symmetrized energy distribution curves (EDC's). These variations are quantitatively represented in Fig. 2c, where the gap values for each band are plotted against photon energy. Remarkably, the superconducting gap of the α band shows a significant reduction from 11 meV (α) / 10 meV ($\alpha_1$) to 8 meV (α) as $k_z$ decreasing from $\pi$ to 0. In contrast, the gaps for the ε and δ bands exhibit comparable but opposite changes: for the ε band, the gap decreases from 11 meV to 8 meV, whereas for the δ band, it increases from 8 meV to 11 meV as $k_z$ varies from $\pi$ to 0. Meanwhile, the superconducting gap of the β band remains unchanged at 7 meV and only a small gap variation for γ band (5 meV to 6 meV) from Z to Γ. We note that the hole bands exhibit no discernible in-plane gap anisotropy [presented in Fig. S3 in Supplementary Materials] which is consistent with the previous ARPES study [37]. Figure 2d shows the symmetrized ARPES spectra along the ZR direction measured at 11.5 K and 40 K, respectively. More data at different temperatures can be found in Fig. S4 in Supplementary Materials. The superconducting gaps disappear at the critical temperature ($T_c \sim 35$ K), demonstrating a BCS-like temperature dependence, as shown in Fig. 2f.

One prominent feature identified in Figs. 1 and 2 is the kink structure in the superconducting phase, which is located precisely on the α band (Fig. S5) and consistently exists in the 3D BZ. Figures 3a-d display a magnified view of the kink along two sets of representative in-plane high symmetry directions (ZA and ZR, as well as ΓM and ΓX), accompanied by the extracted red peak position curves. The white arrows highlight the positions of kinks. The observed kink energies show substantial variation between the $k_z = \pi$ plane (with 24 meV along ZA and 23 meV along ZR below the Fermi level) and the $k_z = 0$ plane, (with 19 meV along both ΓM and ΓX). The self-energy analysis of the corresponding kinks (Figs. 3e-h and Fig. S6) also reveals sharp anomalies at those energies in both the real part (ReΣ) and imaginary part (|ImΣ|) of self-energies. The distinct transition in ImΣ and the peak structure in ReΣ suggest that the associated bosonic mode possesses a relatively narrow energy width. Typically, the presence of a kink in ARPES spectra is indicative of electron coupling with a specific bosonic mode. For superconductors, the energy of the kink-related bosonic mode (Ω) can be derived by subtracting the superconducting gap (Δ) from the kink energy ($E_{kink}$) [24]. Thus, the ~24 meV kink observed in the ZAR plane and 19 meV kink observed in the ΓMX plane may correspond to bosonic modes with energies of 13 meV and 11 meV, respectively.

We further conducted experiments to explore the temperature-dependent evolution of the kink structure in the α band. Figure 4a shows the ARPES spectra along the ZR direction measured at 11.5 K and 40 K, respectively. It is evident that the kink on the α band which is visible at 11.5 K, disappears at 40 K, which exceeds $T_c \sim 35$ K. Figure 4b shows the fitted results of the α band at different temperatures, demonstrating renormalization upon transition into the superconducting phase. Self-energy analysis of the α band at different temperatures can be found in Supplementary Fig. S7, which clearly reveals that both $Re\Sigma$ and $|Im\Sigma|$ exhibit anomalies around 23 meV along ZR direction below $T_C$ while no anomaly in the normal phase. The strength of electron-boson coupling ($\lambda_{e-b}$) can be determined by analyzing the slope change between the bare and renormalized band dispersions, where we select the curve at 40 K as the bare band and $\lambda_{e-b}$ is defined as: $\lambda_{e-b} = (\frac{d\varepsilon}{dk})_{bare}/(\frac{d\varepsilon}{dk})_{renormalized} - 1$. The extracted $\lambda_{e-b}$ values follow a BCS-like temperature dependence as shown in Fig. 4c. Figure 4d further delineates the temperature dependence of $E_{kink}$ and the gap size ($\Delta_\alpha$) of the α band, both of which also follow a BCS-like trend. The bosonic mode energy, calculated as $E_{kink} - \Delta_\alpha$, remains nearly constant below $T_c$, underscoring a robust correlation between the bosonic mode and the superconductivity.

The kink feature in CaKFe$_4$As$_4$ manifests precisely at the onset of superconducting condensation, suggesting that the associated bosonic mode is not phonon, which would typically be observable in both the normal and superconducting phases [40-44]. Instead, a mode only shows up in the superconducting phase of unconventional superconductors is the spin resonance mode. For CaKFe$_4$As$_4$, the kink is located at the energy $E_{kink} = \Delta + E_R$ [5,21-23], where $E_R$ is the energy of the spin resonance mode. Previous INS experiments have resolved two odd resonance modes (10.5 meV and 13 meV) and one even mode (18 meV) [11] in this compound. Correspondingly, our ARPES analysis has detected bosonic mode energies at 11 meV and 13 meV, based on the $k_z$-dependent kink positions. In Fig. 4e, we compare Eliashberg functions ($\alpha^2 F(\omega)$) for these bosonic modes with the momentum-integrated spin resonance curves from INS, showing qualitative consistency, except for the 18 meV even mode with much weaker intensity. Moreover, the resonance energy $E_R$ adheres to an empirical ratio of approximately $E_R/2\Delta \sim 0.64$, commonly among many unconventional superconductors [11,45-48], and the energies of our observed bosons accurately map onto this ratio. This correlation strongly indicates that the bosonic modes we observe in ARPES correspond directly to the spin resonance modes identified in INS.

In terms of spin resonance, antiferromagnetic fluctuations, driven by interband interaction between the hole and electron pockets, dominate the superconducting pairing, resulting in a sign-revered s wave ($s_\pm$) symmetry. This is characterized by an in-plane superconducting gap described by the simple $s_\pm$ gap function $\Delta = \Delta_0 |cosk_x cosk_y|$ [49]. In Fig. 4f, most of the measured gaps along both ZA and ΓM can be well fitted by this function with different $\Delta_0$ values ($\Delta_0 = 15$ meV at $k_z = \pi$ and $\Delta_0 = 12$ meV at $k_z = 0$), confirming the $s_\pm$ symmetry and the antiferromagnetic fluctuations therein. The observed variation in $\Delta_0$ is similar to that in Ba$_{0.6}$K$_{0.4}$Fe$_2$As$_2$ [50], suggesting a significant role for interlayer coupling between FeAs layers in shaping the $k_z$-dependence of the superconducting gaps and kinks. Furthermore, the nearly same Fermi vectors ($k_F$) for α and ε bands further support the presence of $s_\pm$ symmetry through nesting. Interestingly, the same gap size across the α band ($d_{xz}/d_{yz}$) at $k_z = \pi$ ($k_z = 0$), the ε band ($d_{xz}/d_{yz}$) at $k_z = \pi$ ($k_z = 0$) and the δ band ($d_{xy}$) at $k_z = 0$ ($k_z = \pi$) suggest contributions from both interband intraorbital (in-plane) and interband interorbital (out-of-plane) couplings to the pairing mechanism probably in different pairing channels. Previous polarized neutron scattering experiments on CaKFe$_4$As$_4$ have indeed revealed the coexistence of in-plane and out-of-plane spin fluctuations, with a preference for the lower odd mode in the out-of-plane fluctuations and a preference for the larger odd mode in the in-plane fluctuations [11,35,51]. The α band ($3d_{xz}/d_{yz}$) at $k_z = 0$ hybridizes more with $3d_z^2/4p_z$ than the α band at $k_z = \pi$ due to the topological band inversion [36]. Therefore, the lower mode energy at $k_z = 0$ might be due to the greater contribution of out-of-plane fluctuations than in-plane ones.

**CONCLUSION**

Our findings elucidate clear electron-spin fluctuation coupling in CaKFe$_4$As$_4$, strongly associated with its superconductivity. This highlights the essential role of spin resonance in Fe-based superconductors. The interlayer coupling in this bilayer FeAs compound facilitates a three-dimensional pairing between the hole and electron pockets. Given the ubiquitous presence of spin resonance across various Fe-based superconductors, our findings potentially suggest a widespread phenomenon of kink induced by electron and spin exciton coupling, thereby advancing our understanding of superconductivity in this class of materials.

**METHODS**

High quality CaKFe$_4$As$_4$ single crystals were synthesized by the solid-state reaction method described in Ref. [34]. Synchrotron-ARPES measurements were carried out using synchrotron light sources at BL 03U of Shanghai Synchrotron Radiation Facility (SSRF) in China. The overall energy resolution was set to be better than 6 meV at 30 eV photon energy and the angular resolution is ~ 0.2 degree for the gap and kink measurements. The crystals were cleaved *in-situ* and measured with a base pressure better than $6 \times 10^{-11}$ Torr. All the data presented in this paper were taken within a few hours after cleavage ensuring the results were not affected by aging effect.

**ACKNOWLEDGEMENTS**

This work is supported by the National Key R&D Program of China (Grant No. 2023YFA1406304 (J. J), the National Natural Science Foundation of China (Grant No. 12174362 (J. J), No. 11790312 (D. L. F), No. 11888101 (D. L. F), No. 92065202 (J. J)), the Innovation Program for Quantum Science and Technology (No. 2021ZD0302803 (D. L. F) and the New Cornerstone Science Foundation (D. L. F). Part of this research used Beamline 03U of the Shanghai Synchrotron Radiation Facility, which is supported by ME2 project under contract no. 11227902 from National Natural Science Foundation of China.

**COMPETING INTERESTS**

The authors declare no competing interest.

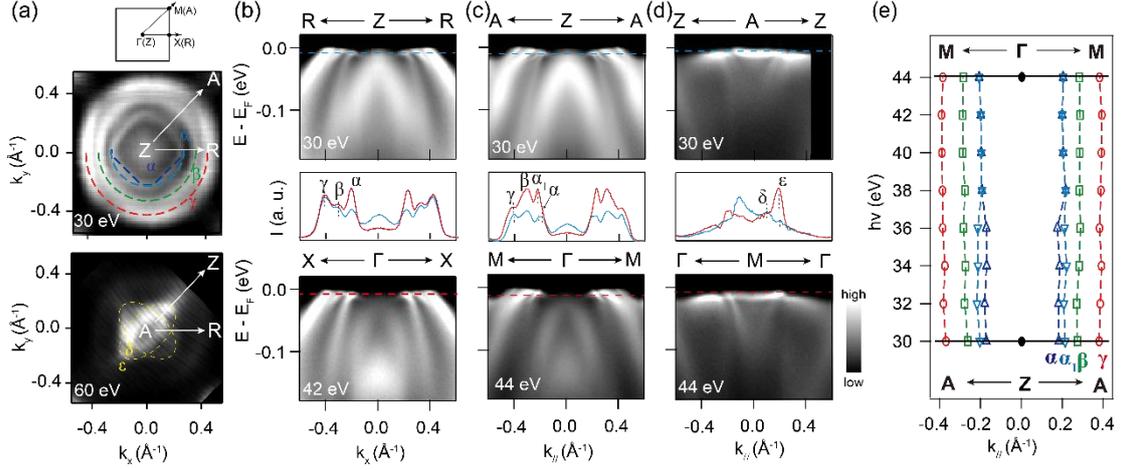

FIG. 1. Basic electronic structure of CaKFe$_4$As$_4$. (a) Fermi surface maps in ZAR plane ($k_z = \pi$) obtained at 11 K using 30 eV and 60 eV photons, superposed with high symmetry directions. The deep and light blue dashed lines are the indication of the two splitting bands of α, the green dashed line refers to β and the red dashed line refer to γ. Two electron pockets ε and δ are located around the A point as indicated by the overlapped yellow dashed ellipses. (b) ARPES spectra of the hole bands along RZR ($k_z = \pi$, using 30 eV photons) and XΓX ($k_z = 0$, using 42 eV photons) in the upper and lower parts, respectively. The middle part shows the extracted momentum distribution curves (MDC) at the blue and red dashed lines. (c) ARPES spectra and corresponding MDC of the hole bands along AZA and MΓM. (d) ARPES spectra and corresponding MDC of the electron bands along ZAZ and ΓMΓ. (e) The extracted peak positions of the hole bands at various photon energies. It shows no observable dispersion in the $k_z$ direction, except for the splitting behavior of the α band.

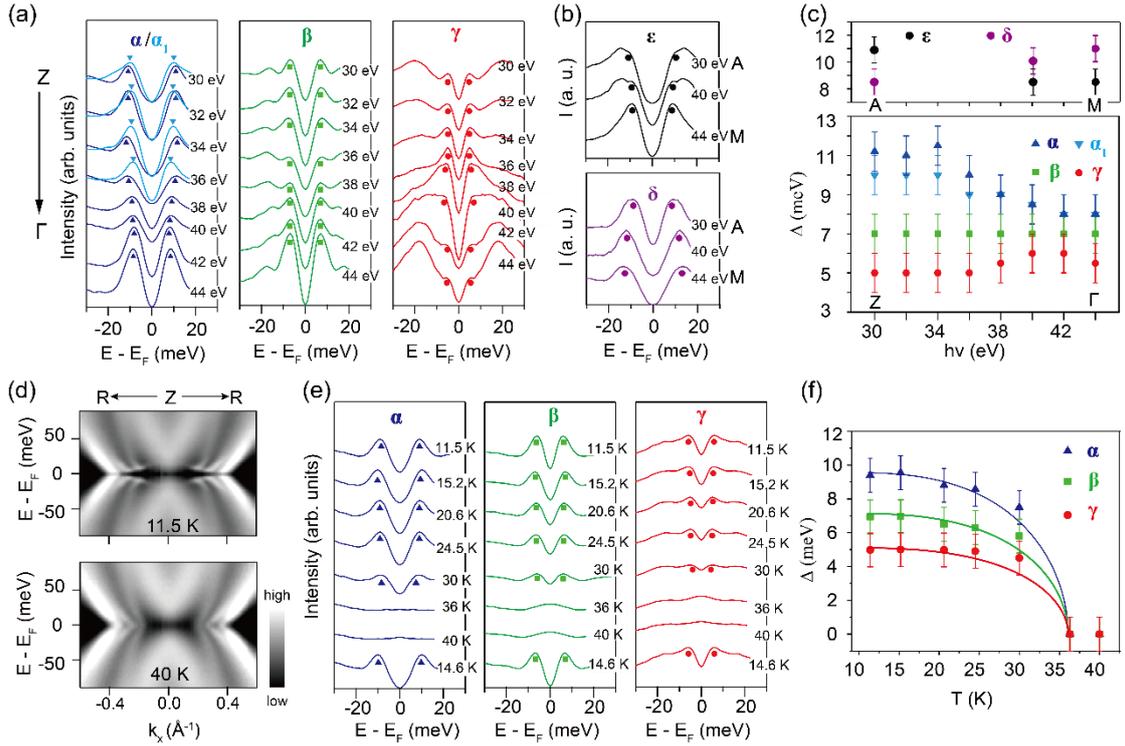

FIG. 2. The $k_z$ dependence and temperature evolution of superconducting gaps. (a) Symmetrized EDCs the α, β and γ bands at various photon energies (from 30 eV to 44 eV) with intensity shifts for clarity. The splitting band, $α_1$, exists near the Z point (30 eV), and degenerates with α as it approaches Γ (44 eV). (b) Symmetrized EDCs the electron bands, ε and δ, from the A point (30 eV) to the M point (44eV). (c) Extracted superconducting gaps as a function of photon energy. (d) Symmetrized ARPES spectra along ZR below and above Tc. (e) Symmetrized EDCs of the α, β and γ bands at various temperatures. The superconducting gaps close above $T_c$ and recover upon cooling. (f) Superconducting gap values of the α, β and γ bands as a function of temperature. The gaps are well fitted with the solid lines predicted by the BCS theory.

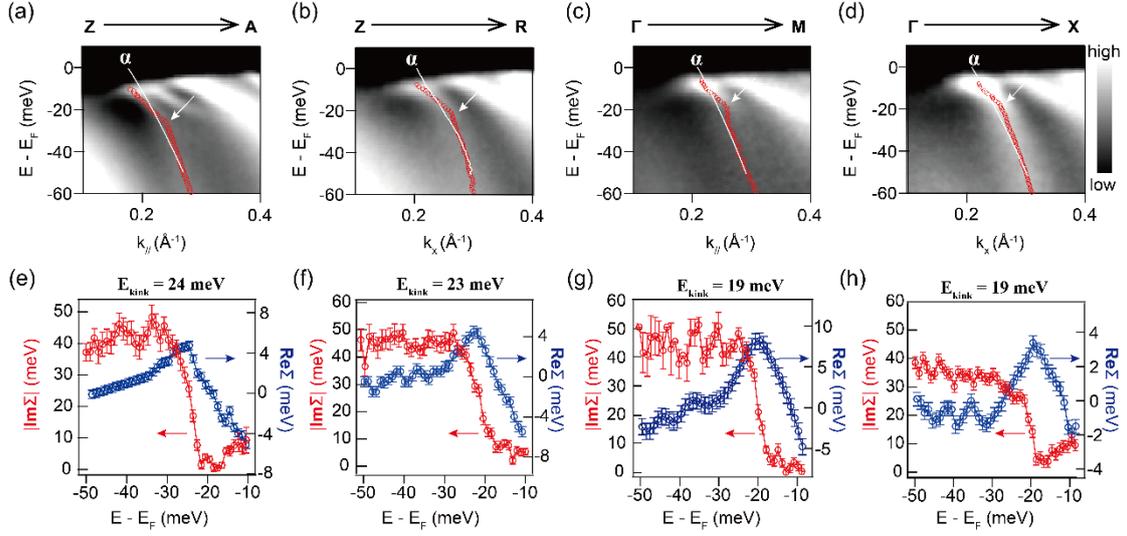

FIG. 3. The $k_z$ dependence of the kinks. (a,b) The kink structures along ZA and ZR in the ZAR plane, overlapped with the red peak position curves. The white lines represent the bare band dispersion of the α band, determined by the same Fermi vector ($k_F$) and near closely overlap with the red curves at higher binding energies. (c,d) The kink structure along ΓM and ΓX in the ΓMX plane. The white arrows indicate the kink positions. (e,f) Self-energy analysis of the kink in (a,b). The real (ReΣ) and imaginary parts (|ImΣ|) of self-energy show sharp anomalies are located at 24 meV and 23 meV below the Fermi level along ZA and ZR, respectively. (g,h) Self-energy analysis of the kink in (c,d). The unusual upturn near the Fermi level in (h) is due to the artifacts induced by Bogliubov bending band of the nearby β band. The anomalies located 19 meV below the Fermi level are resolved by the ReΣ and |ImΣ| curves of the kinks.

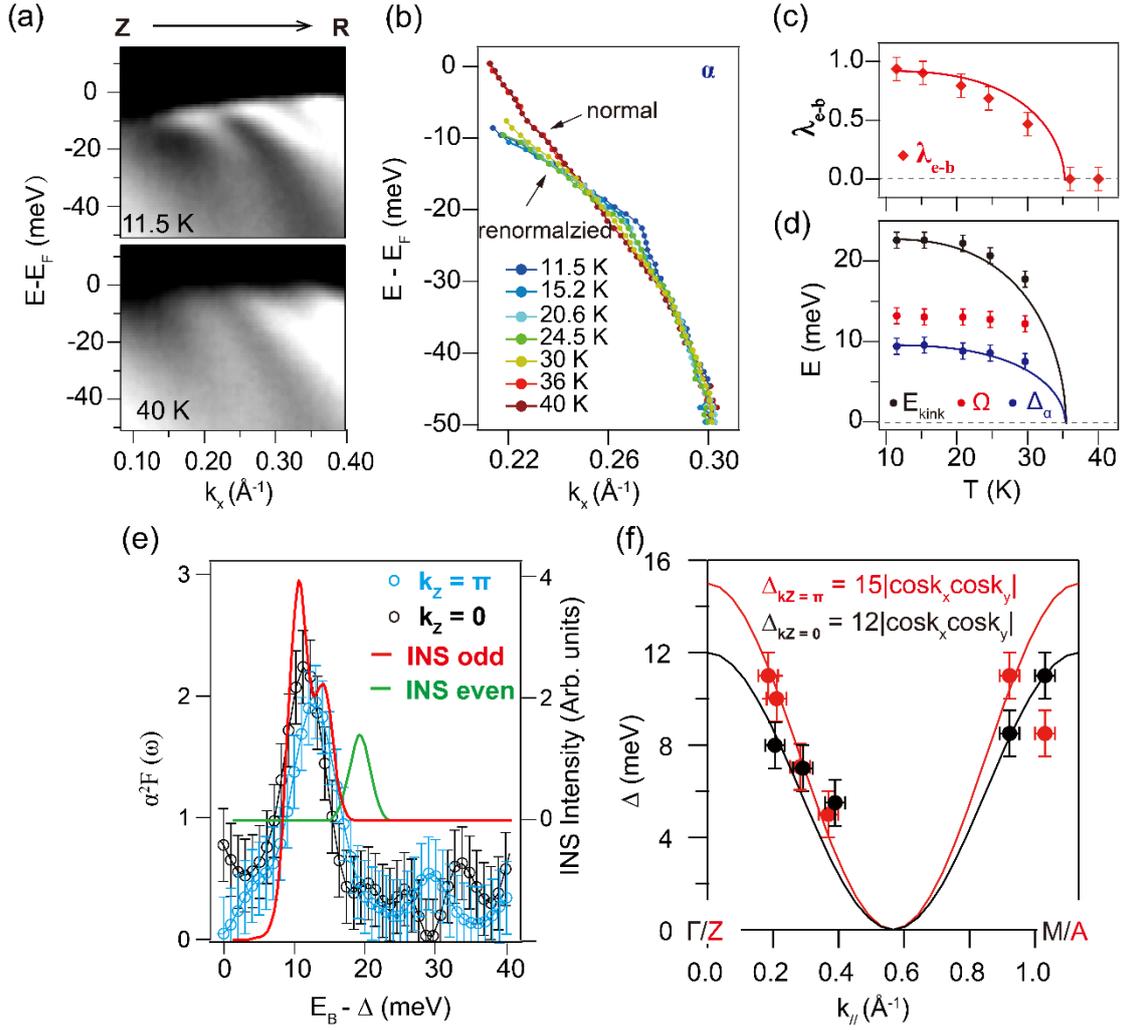

FIG. 4. Evidence of the observation of spin exciton. (a) ARPES spectrum along the ZR direction measured at 11.5 K and 40 K. (b) Extracted band dispersion of the α band at various temperatures. The kink structure gradually weakens and eventually dissipates as temperature surpasses $T_c$. The curve of 40 K is viewed as the normal dispersion. (c) Electron-boson coupling strength, $\lambda_{e-b}$, as a function of temperature. The red solid curve is the fitting results by the BCS theory. (d) The kink energy ($E_{kink}$), $\Delta_\alpha$ and the extracted bosonic mode energy ($\Omega = E_{kink} - \Delta_\alpha$) as a function of temperature. Both $E_{kink}$ and $\Delta_\alpha$ follow the BCS-like temperature dependence, while the $\Omega$ only has a slight change. (e) Eliashberg functions ($\alpha^2 F(\omega)$) of the α band at $k_z = \pi$ and $k_z = 0$ compared with the spin resonance curve in INS results. The red and green curves are two odd modes (peaks at 10.5 meV and 13 meV in the red curve) extracted from a previous neutron study [11]. (f) In-plane distributions of the superconducting gaps. Most of the in-plane superconducting gaps at $k_z = \pi$ and $k_z = 0$ could be described by simple $s_\pm$ gap function $\Delta = \Delta_0|cosk_x cosk_y|$, where $\Delta_0$ if fitted with 15 meV at $k_z = \pi$ and 12 meV at $k_z = 0$, respectively, while the γ band at $k_z = 0$ and the δ band at $k_z = \pi$ are not well fitted.

**References**

[1]  J. Paglione & R. L. Greene, *High-temperature superconductivity in iron-based materials.* Nat. Phys. **6**, 645-658 (2010).

[2]  R. M. Fernandes, A. V. Chubukov, & J. Schmalian, *What drives nematic order in iron-based superconductors?* Nat. Phys. **10**, 97-104 (2014).

[3]  Q. Si, R. Yu & E. Abrahams, *High-temperature superconductivity in iron pnictides and chalcogenides.* Nat. Rev. Mater. **1**, 4 (2016).

[4]  F. Wang, D.-H. Lee, *The Electron-Pairing Mechanism of Iron-Based Superconductors.* Science **332**, 200-204 (2011).

[5]  M. Eschrig, *The effect of collective spin-1 excitations on electronic spectra in high-$T_c$ superconductors.* Adv. Phys. **55**, 47-183 (2006).

[6]  H. Kontani, R. Tazai, Y. Yamakawa & S. Onari, *Unconventional density waves and superconductivities in Fe-based superconductors and other strongly correlated electron systems.* Adv. Phys. **70**, 355-443 (2023).

[7]  P. Wiecki, M. Frachet, A.-A. Haghighirad, T. Wolf, C. Meingast, R. Heid & A. E. Böhmer, *Emerging symmetric strain response and weakening nematic fluctuations in strongly hole-doped iron-based superconductors.* Nat. Commun. **12(1)**, 4824 (2021).

[8]  H. F. Fong, P. Bourges, Y. Sidis, L. P. Regnault, A. Ivanov, G. D. Gu, N. Koshizuka, B. Keimer, *Neutron scattering from magnetic excitations in $Bi_2Sr_2CaCu_2O_{8+\delta}$.* Nature **398,** 588-591 (1999).

[9]  A. D. Christianson, E. A. Goremychkin, R. Osborn, S. Rosenkranz, M. D. Lumsden, C. D. Malliakas, I. S. Todorov, H. Claus, D. Y. Chung, M. G. Kanatzidis, et al. *Unconventional superconductivity in $Ba_{(0.6)}K_{(0.4)}Fe_2As_2$ from inelastic neutron scattering.* Nature **456**, 930-932 (2008).

[10] D. S. Inosov, J. T. Park, P. Bourges, D. L. Sun, Y. Sidis, A. Schneidewind, K. Hradil, D. Haug, C. T. Lin, B. Keimer, et al. *Normal-state spin dynamics and temperature-dependent spin-resonance energy in optimally doped $BaFe_{1.85}Co_{0.15}As_2$.* Nat. Phys. **6**, 178-181 (2009).

[11] T. Xie, Y. Wei, D. L. Gong, T. Fennell, U. Stuhr, R. Kajimoto, K. Ikeuchi, S. L. Li, J. P. Hu and H. Q. Luo, *Odd and Even Modes of Neutron Spin Resonance in the Bilayer Iron-Based Superconductor $CaKFe_4As_4$.* Phys. Rev. Lett. **120**, 267003 (2018).

[12] Q. Wang, J. T. Park, Y. Feng, Y. Hao, B. Pan, J. W. Lynn, A. Ivanov, S. Chi, M. Matsuda, H. Cao, et al. *Transition from Sign-Reversed to Sign-Preserved Cooper-Pairing Symmetry in Sulfur-Doped Iron Selenide Superconductors.* Phys. Rev. Lett. **116**, 197004 (2016).

[13] K. Iida, M. Ishikado, Y. Nagai, H. Yoshida, A. D. Christianson, N. Mura, K. Kawashim, Y. Yoshida, H. Eisaki, and A. Iyo, *Spin Resonance in the New-Structure-Type Iron-Based Superconductor $CaKFe_4As_4$.* Journal. Phys. Soc. Jpn. **86**, 093703 (2017).

[14] J. Zhao, L. P. Regnault, C. Zhang, M. Wang, Z. Li, F. Zhou, Z. Zhao, C. Fang, J. Hu, and P. Dai, *Neutron spin resonance as a probe of the superconducting energy gap of $BaFe_{1.9}Ni_{0.1}As_2$ superconductors.* Phys. Rev. B **81**, 180505(R) (2010).

[15] N. Qureshi, C. H. Lee, K. Kihou, K. Schmalzl, P. Steffens & M. Braden, *Anisotropy of incommensurate magnetic excitations in slightly overdoped $Ba_{0.5}K_{0.5}Fe_2As_2$ probed by polarized inelastic neutron scattering experiments.* Phys. Rev. B **90**, 100502(R) (2014).

[16] L. Capogna, B. Fauqué, Y. Sidis, C. Ulrich, P. Bourges, S. Pailhès, A. Ivanov, J. L. Tallon, B.

Liang, C. T. Lin, et al. *Odd and even magnetic resonant modes in highly overdoped $Bi_2Sr_2CaCu_2O_{8+\delta}$.* Phys. Rev. B **75**, 060502(R) (2007).

[17] S. Pailhès, C. Ulrich, B. Fauqué, V. Hinkov, Y. Sidis, A. Ivanov, C. T. Lin, B. Keimer and P. Bourges, *Doping dependence of bilayer resonant spin excitations in (Y, Ca)$Ba_2Cu_3O_{6+x}$.* Phys. Rev. Lett. **96**, 257001 (2006).

[18] M. D. Lumsden, A. D. Christianson, D. Parshall, M. B. Stone, S. E. Nagler, G. J. MacDougall, et al. *Two-dimensional resonant magnetic excitation in $BaFe_{1.84}Co_{0.16}As_2$.* Phys. Rev. Lett. **102**, 107005 (2009).

[19] S. Chi, A. Schneidewind, J. Zhao, L. W. Harriger, L. Li, Y. Luo, G. Cao, Z. Xu, M. Loewenhaupt, J. Hu, and P. Dai, *Inelastic neutron-scattering measurements of a three-dimensional spin resonance in the FeAs-based $BaFe_{1.9}Ni_{0.1}As_2$ superconductor.* Phys. Rev. Lett. **102**, 107006 (2009).

[20] J. P. Castellan, S. Rosenkranz, E. A. Goremychkin, D. Y. Chung, I. S. Todorov, M. G. Kanatzidis, et al. *Effect of Fermi surface nesting on resonant spin excitations in $Ba_{(1-x)}K_{(x)}Fe_2As_2$.* Phys. Rev. Lett. **107**, 177003 (2011).

[21] A. Kaminski, M. Randeria, J. C. Campuzano, M. R. Norman, H. Fretwell, J. Mesot, T. Sato, T. Takahashi and K. Kadowaki, *Renormalization of Spectral Line Shape and Dispersion below Tc in $Bi_2Sr_2CaCu_2O_{8+\delta}$.* Phys. Rev. Lett. **86**, 1070 (2001).

[22] A. D. Gromko, A. V. Fedorov, Y. D. Chuang, J. D. Koralek, Y. Aiura, Y. Yamaguchi, K. Oka, Y. Ando and D. S. Dessau, *Mass-renormalized electronic excitations at „□,0… in the superconducting state of $Bi_2Sr_2CaCu_2O_{8+\delta}$.* Phys. Rev. B **68**, 174520 (2003).

[23] A. Abanov and A. V. Chubukov, *A Relation between the Resonance Neutron Peak and ARPES Data in Cuprates.* Phys. Rev. Lett. **83, 1652** (1999).

[24] P. Richard, T. Sato, K. Nakayama, S. Souma, T. Takahashi, Y.-M. Xu, G. F. Chen, J. L. Luo, N. L. Wang, and H. Ding *Angle-resolved photoemission spectroscopy of the Fe-Based $Ba_{0.6}K_{0.4}Fe_2As_2$ high temperature superconductor: evidence for an orbital selective electron-mode coupling.* Phys. Rev. Lett. **102**, 047003 (2009).

[25] H. Iwasawa, J. F. Douglas, K. Sato, T. Masui, Y. Yoshida, Z. Sun, H. Eisaki, H. Bando, A. Ino, M. Arita, et al. *Isotopic Fingerprint of Electron-Phonon Coupling in High-Tc Cuprates.* Phys. Rev. Lett. **101**, 157005 (2008).

[26] F. Mazzola, J. W. Wells, R. Yakimova, S. Ulstrup, J. A. Miwa, R. Balog, M. Bianchi, M. Leandersson, J. Adell, P. Hofmann, and T. Balasubramanian. *Kinks in the σ Band of Graphene Induced by Electron-Phonon Coupling.* Phys. Rev. Lett. **111**, 216806 (2013).

[27] Y. Zhong, S. Li, H. Liu, Y. Dong, K. Aido, Y. Arai, H. Li, W. Zhang, Y. Shi, Z. Wang, et al. *Testing electron–phonon coupling for the superconductivity in kagome metal $CsV_3Sb_5$.* Nat. Commun. **14**, 1945 (2023).

[28] Y. Wu, W. Zhang, Y. Fang, S. Lu, L.Wang, P. Li, Z. Wu, Z. Xiao, C. Cao, X. Wang, et al. *Interfacial electron-phonon coupling and quantum confinement in ultrathin Yb films on graphite.* Phy. Rev. B **104**, L161402 (2021).

[29] J. S. Zhou, R. Z. Xu, X. Q. Yu, F. J. Cheng, W. X. Zhao, X. Du, S. Z. Wang, Q. Q. Zhang, X. Gu, S. M. He, et al. *Evidence for Band Renormalizations in Strong-Coupling Superconducting Alkali-Fulleride Films.* Phys. Rev. Lett. **130**, 216004 (2023).

[30] T. L. Yu, M. Xu, W. T. Yang, Y. H. Song, C. H. P. Wen, Q. Yao, T. Zhang, W. Li, X. Y. Wei, et al. *Strong band renormalization and emergent ferromagnetism induced by electron-*

*antiferromagnetic-magnon coupling.* Nat. Commun. **13**, 6560 (2022).

[31] P. Li, S. Liao, Z. Wang, H. Li, S. Su, J. Zhang, Z. Chen, Z. Jiang, L. Huai, J. He, et al. *Evidence of electron interaction with an unidentified bosonic mode in superconductor $CsCa_2Fe_4As_4F_2$.* Nat. Commun. **15**, 6433 (2024).

[32] L. Wray, D. Qian, D. Hsieh, Y. Xia, L. Li, J. G. Checkelsky, A. Pasupathy, K. K. Gomes, C. V. Parker, A. V. Fedorov, et al. *Momentum dependence of superconducting gap, strong-coupling dispersion kink, and tightly bound Cooper pairs in the high-Tc $(Sr,Ba)_{1-x}(K,Na)_xFe_2As_2$ superconductors.* Phys. Rev. B **78**, 184508 (2008).

[33] W. R. Meier, T. Kong, U. S. Kaluarachchi, V. Taufour, N. H. Jo, G. Drachuck, A. E. Böhmer, S. M. Saunders, A. Sapkota, A. Kreyssig, et al. *Anisotropic thermodynamic and transport properties of single-crystalline $CaKFe_4As_4$.* Phys. Rev. B **94**, 064501 (2016).

[34] A. Iyo, K. Kawashima, T. Kinjo, T. Nishio, S. Ishida, H. Fujihisa, Y. Gotoh, K. Kihou, H. Eisaki, and Y. Yoshida, *New-Structure-Type Fe-Based Superconductors: $CaAFe_4As_4$ (A = K, Rb, Cs) and $SrAFe_4As_4$ (A = Rb, Cs).* J. Am. Chem. Soc. **138**, 3410-3415 (2016).

[35] T. Xie, C. Liu, F. Bourdarot, L.-P. Regnault, S. Li and H. Luo, Spin-excitation anisotropy in the bilayer iron-based superconductor $CaKFe_4As_4$. *Phys. Rev. Research* **2**, 022018(R) (2020).

[36] W. Liu, L. Cao, S. Zhu, L. Kong, G. Wang, M. Papaj, P. Zhang, Y. Liu, H. Chen, G. Li, et al. *A new Majorana platform in an Fe-As bilayer superconductor.* Nat. Commun. **11**, 5688 (2020).

[37] D. Mou, T. Kong, W. R. Meier, F. Lochner, L. Wang, Q. Lin, Y. Wu, S. L. Bud'ko, I. Eremin, D. D. Johnson, P. C. Canfield, and A, Kaminski, *Enhancement of the Superconducting Gap by Nesting in $CaKFe_4As_4$: A New High Temperature Superconductor.* Phys. Rev. Lett. **117**, 277001 (2016).

[38] X. Yu, Z. Wei, Z. Zhao, T. Xie, C. Liu, G. He, Q. Chen, L. Shan, H. Luo, Q. Huan, J. Yuan, and K. Jin, *Surface morphology and electronic structure in stoichiometric superconductor $CaKFe_4As_4$ probed by scanning tunneling microscopy/spectroscopy.* Sci. China-Phys. Mech. Astron. **64**, 127411 (2021).

[39] W. R. Meier, T. Kong, S. L. Budko and P. C. Canfield, *Optimization of the crystal growth of the superconductor $CaKFe_4As_4$ from solution in the $FeAs-CaFe_2As_2-KFe_2As_2$ system.* Phys. Rev. Mater. **1**, 013401 (2017).

[40] W. L. Zhang, W. R. Meier, T. Kong, P. C. Canfield and G. Blumberg, *High-Tc superconductivity in $CaKFe_4As_4$ in absence of nematic fluctuations.* Phys. Rev. B **98**, 140501(R) (2018).

[41] R. Yang, Y. Dai, B. Xu, W. Zhang, Z. Qiu, Q. Sui, C. C. Homes and X. Qiu, *Anomalous phonon behavior in superconducting $CaKFe_4As_4$: An optical study.* Phys. Rev. B **95**, 064506 (2017).

[42] D. Jost, J. R. Scholz, U. Zweck, W. R. Meier, A. E. Böhmer, P. C. Canfield, N. Lazarević and R. Hackl, *Indication of subdominant d-wave interaction in superconducting $CaKFe_4As_4$.* Phys. Rev. B **98**, 020504(R) (2018).

[43] F. Stramaglia, G. M. Pugliese, L. Tortora, L. Simonelli, C. Marini, W. Olszewski, S. Ishida, A. Iyo, H. Eisaki, T. Mizokawa, and N. L. Saini, *Temperature Dependence of the Local Structure and Iron Magnetic Moment in the Self-Doped $CaKFe_4As_4$ Iron-Based Superconductor.* J. Phys. Chem. C **125**, 10810 (2021).

[44] H. Zhao, R. Blackwell1, M. Thinel, T. Handa, S. Ishida, X. Zhu, A. Iyo, H. Eisaki, A. N. Pasupathy, & K. Fujita, *Smectic pair-density-wave order in $EuRbFe_4As_4$.* Nature **618**, 940-945 (2023).

[45] W. Hong, L. Song, B. Liu, Z. Li, Z. Zeng, Y. Li, D. Wu, Q. Sui, T. Xie, S. Danilkin, et al. *Neutron*

| | |
|---|---|
| | *Spin Resonance in a Quasi-Two-Dimensional Iron-Based Superconductor.* Phys. Rev. Lett. **125**, 117002 (2020). |
| [46] | Z. Wang, H. Yang, D. Fang, B. Shen, Q. Wang, L. Shan, C. Zhang, P. Dai and H.-H. Wen, *Close relationship between superconductivity and the bosonic mode in $Ba_{0.6}K_{0.4}Fe_2As_2$ and $Na(Fe_{0.975}Co_{0.025})As$.* Nat. Phys. **9**, 42-48 (2012). |
| [47] | G. Yu, Y. Li, E. M. Motoyama & M. Greven, *A universal relationship between magnetic resonance and superconducting gap in unconventional superconductors.* Nat. Phys. **5**, 873-875 (2009). |
| [48] | D. S. Inosov, J. T. Park, A. Charnukha, Y. Li, A. V. Boris, B. Keimer and V. Hinkov, *Crossover from weak to strong pairing in unconventional superconductors.* Phys. Rev. B **83**, 214520 (2011). |
| [49] | J. Hu & H. Ding, *Local antiferromagnetic exchange and collaborative Fermi surface as key ingredients of high temperature superconductors.* Sci. Rep. **2**, 381 (2012). |
| [50] | Y. Xu, Y. Huang, X. Cui, E. Razzoli, M. Radovic, M. Shi, G. Chen, P. Zheng, N. Wang, C. Zhang, et al. *Observation of a ubiquitous three-dimensional superconducting gap function in optimally doped $Ba_{0.6}K_{0.4}Fe_2As_2$.* Nat. Phys. **7**, 198-202 (2011). |
| [51] | C. Liu, P. Bourges, Y. Sidis, T. Xie, G. He, F. Bourdarot, S. Danilkin, H. Ghosh, S. Ghosh, X. Ma, et al. *Preferred Spin Excitations in the Bilayer Iron-Based Superconductor $CaK(Fe_{0.96}Ni_{0.04})_4As_4$ with Spin-Vortex Crystal Order.* Phys. Rev. Lett. **128**, 137003 (2022). |